# Modification of titanium and titanium dioxide surfaces by ion implantation: combined XPS and DFT study

D.W. Boukhvalov,[*,1,2] D.M. Korotin,[3] A.I. Efremov,[3] E.Z. Kurmaev,[3] Ch. Borchers,[4] I.S. Zhidkov,[5] D.V. Gunderov,[6] R.Z. Valiev,[6] N.V. Gavrilov,[7] S.O. Cholakh[5]

[1] Department of Chemistry, Hanyang University, 17 Haengdang-dong, Seongdong-gu, Seoul 133-791, Korea,
E-mail: danil@hanyang.ac.kr
[2] School of Computational Sciences, Korea Institute for Advanced Study (KIAS), Hoegiro 87, Dongdaemun-Gu, Seoul, 130-722, Korea
[3] Institute of Metal Physics, Russian Academy of Sciences-Ural Division, S. Kovalevskoi Street 18, 620990 Yekaterinburg, Russia
[4] Institute for Material Physics, University of Göttingen, Friedrich-Hund-Platz 1, 37077 Goettingen, Germany
[5] Ural Federal University, Mira Street 9, 620002 Yekaterinburg, Russia
[6] Institute of Physics of Advanced Materials, Ufa State Aviation Technical University, Ufa 450000, Russia
[7] Institute of Electrophysics, Russian Academy of Sciences-Ural Division, 620016 Yekaterinburg, Russia

*The results of XPS measurements (core levels and valence bands) of $P^+$, $Ca^+$, $P^+Ca^+$ and $Ca^+P^+$ ion implanted (E=30 keV, D=1x10$^{17}$ cm$^{-2}$) commercially pure titanium (cp-Ti) and first-principles density functional theory (DFT) calculations demonstrates formation of various structural defects in titanium dioxide films formed on the surface of implanted materials. We have found that for double implantation (Ti:$P^+$,$Ca^+$ and Ti:$Ca^+$,$P^+$) the outermost surface layer formed mainly by Ca and P, respectively, i.e. the implantation sequence is very important. The DFT calculations show that under $P^+$ and $Ca^+P^+$ ion implantation the formation energies for both cation (P-Ti) and anion (P-O) substitutions are comparable which can induce the creation of $[PO_4]^{3-}$ and Ti-P species. For $Ca^+$ and $P^+Ca^+$-ion implantation the calculated formation energies correspond to $Ca^{2+}$-$Ti^{4+}$ cation substitution. This conclusion is in agreement with XPS Ca 2p and Ti 2p core levels and valence band measurements and DFT calculations of electronic structure of related compounds. The conversion of implanted ions to $Ca^{2+}$ and $[PO_4]^{3-}$ species provides a good biocompatibility of cp-Ti for further formation of hydroxyapatite.*

**1 Introduction** The materials used as medical implants should have an optimal combination of good mechanical strength with high chemical stability, corrosion resistance and biocompatibility [1]. One of such materials is commercially pure titanium (*cp*-Ti) which displays the mechanical properties similar to bone, has also all other above mentioned properties and for this reason widely used as dental and orthopedic implant as well as for osteosynthesis applications [2]. These remarkable properties of *cp*-Ti are due to existence of protective stable native film of $TiO_2$ (2-20 nm) which provides the corrosion resistance, can adsorb proteins and induce the differentiation of bone cells [3]. However, titanium and bones are generally separated by a thin non-mineral layer [4] and true adhesion of titanium to bones has not been observed. In order to make titanium biologically bond to bones, surface modification methods have been proposed to improve the bone conductivity or bioactivity of titanium. A common way to improve the osseointegration is the coating of titanium surface with hydroxyapatite (HA; the mineral component of the bone). Many processes like plasma-spraying, electrolytic oxidation, sol-gel, sputtering, ion implantation, laser cladding, etc. are used to produce HA coatings on implant metal surfaces [5]. Among these methods the ion implantation seems to be very attractive and preferable [6] because HA coatings obtained by plasma-spraying or sputtering are too thick and can be easily broken within the thick film and in the interface between the film and substrate [7-8].

The electrolytic processes produce coatings with poor adhesion strengths and sol-gel processes involve longer times and require a post-processing annealing to obtain more desirable properties [6].

Ion implantation gives a possibility to inject any element into the near-surface region of any substrate [9-10]. This is a non-equilibrium process and provides producing materials with compositions and structures which can not be obtained with help of conventional equilibrium methods such as thermal diffusion or alloying. Ion implantation is performed in vacuum and therefore it is an ultra-clean process and by such a way high purity layers can be prepared. There is no strict limit on the solubility due to a non-equilibrium process and hence solid solubility limit can be exceeded (see Ref. 6).

In the given paper we present the results of characterization of $Ca^+$ and/or $P^+$ ion implanted *cp*-Ti by means of XPS technique and by comparison of XPS measurements with specially performed first-principles electronic structure calculations of related compounds. There are a lot of previous XPS measurements of Ca and P ion implanted titanium (see [11-18] and references therein) but as a rule they are limited by the measurements only of core level spectra. We have performed at first time the high-energy resolved XPS measurements of not only core levels but also valence bands (XPS VB) of $Ca^+$ or/and $P^+$ ion implanted *cp*-Ti. The comparison of XPS VB with specially performed electronic structure calculations of reference compounds such as, TiP and $H_3PO_4$ provided a full characterization of electronic structure of ion implanted *cp*-Ti and allowed to estimate the biocompatibility of near the surface layers.

## 2 Experimental and theoretical details

Commercially pure titanium *cp*-Ti Grade 4 (Ti: base, C: 0.052%, O: 0.34%, Fe: 0.3%, N: 0.015% (wt.%)) was employed to conduct the research. The coarse grained *cp*-Ti in the form of discs (10 mm dia and 1 mm thick) was annealed at $950^0C$ for 1 hr. Then the surface of the substrates ware grinded using of diamond suspension with grain size less than 1 μm at the finishing stage, after the substrates were cleaned in ultrasonic bath in acetone solution. The implantation of $P^+$ or/and $Ca^+$ ions in *cp*-T was carried out in oil-free vacuum provided by a turbomolecular pump. Before irradiation the vacuum chamber was evacuated to a residual pressure of $3\times10^{-3}$ Pa. A $Ca^+$ ion beam with energy of 30 keV was generated at the MEVVA-type ion source based on pulsed arc with a calcium cathode. Auxiliary discharge in argon is used for arc ignition, so the gas pressure in the chamber during irradiation increased to $1.5\times10^{-2}$ Pa. The processing was carried out in a pulsed mode with a repetition rate of 12.5 Hz with a current pulse duration of 0.4 ms, pulse current density was 4 mA/cm². The duration of exposure for which the ion fluence reached $1\times10^{17}$ cm$^{-2}$, was 13 minutes 20 seconds. A source of phosphorus ion beam was based on a low-pressure (0.01 Pa) glow discharge with a hollow cathode in magnetic field (3.6 mT). At first the discharge in argon (0.5 A, 800 V) heats containers filled with red phosphorus till the temperature (~300°C) at which phosphorus partial pressure reaches ~0.01 Pa. Then argon feeding was stopped, and the discharge (0.2 A, 800 - 1000 V) operates in phosphorus vapours. The ion source operates at dc mode with accelerating voltage equal to 25 kV and the ion current density 0.07 mA/cm$^2$. The ion fluency reached the value of $1 \times 10^{17}$ cm$^{-2}$ during 3 minutes 50 seconds. During the implantation the samples were placed on a massive water-cooled collector. The initial temperature of the samples prior to irradiation was 20°C. After implantation, the samples were cooled in a vacuum for 20 min.

XPS core-level and valence-band spectra measurements were made using PHI XPS Versaprobe 5000 spectrometer (ULVAC-Physical Electronics, USA, started in the year of 2011). Versaprobe 5000 is based on the classic x-ray optic scheme with hemispherical quartz monochromator and energy analyzer working in the range of binding energies from 0 to 1500 eV. The most advantage of this XPS system is an electrostatic focusing and magnetic screening. As a result, the achieved energy resolution is $\Delta E \leq 0.5$ eV for Al *Kα* excitation (1486.6 eV). The vacuum circuit of Analytical Chamber uses oil-free rotary pump which allow to obtain and keep the pressure not less than $10^{-7}$ Pa. The dual-channel neutralizer (PHI patent) was applied in order to compensate the local charging of the sample under study due to the loss of photoelectrons during XPS measurements. All samples under study were previously kept in the Intro Chamber within 24 hours under rotary pumping. After that the sample was introduced into the Analytical Chamber and controlled with the help of "Chemical State Mapping" mode in order to detect micro impurities. If the micro impurities were detected then the sample was replaced from the reserved batch.

The XPS spectra of core-levels and valence band were recorded in Al $K\alpha$ 100 μm spot mode with x-ray power load of the sample less than 25 Watts. Typical signal to noise ratio values in this case were not less than 10000/3. The spectra were processed using ULVAC-PHI MultiPak Software v. 9.3 and the residual background was removed with the help of Tougaard method [19]. XPS spectra were calibrated using the reference energy value of carbon core-level $E$ (C $1s$) = 285.0 eV.

X-ray diffraction (XRD) patterns of pristine and Ca and/or P implanted *cp*-Ti were measured with a DRON-6 X-ray diffractometer with a monochromatized Cr - K$\alpha$ radiation.

The density-functional theory (DFT) calculations of formation energies for different configurations of structural defects were performed using the SIESTA pseudopotential code [20], as utilized with success previously for related studies of impurities pairs in semiconductors [21]. All calculations were performed using the Perdew-Burke-Ernzerhof variant of the generalized gradient approximation (GGA-PBE) [22] for the exchange-correlation potential. A full optimization of the atomic positions was performed during which the electronic ground state was consistently found using norm-conserving pseudopotentials for cores and a double-$\zeta$ plus polarization basis of localized orbitals for Ca, Ti, P, O and H. Optimizations of the force and total energy were performed with an accuracy of 0.04 eV/Å and 1 meV, respectively.

Because the surface of *cp*-Ti is covered by $TiO_2$ oxide layer we analyzed the formation of different structural defects for $TiO_2$:$P^+$, $TiO_2$:$Ca^+$, $TiO_2$:$P^+$,$Ca^+$ and $TiO_2$:$Ca^+$,$P^+$. We have used $TiO_2$ in anatase phase with supercell consisting of 96 atoms within three dimensional periodic boundary conditions as the model of bulk titanium dioxide and within two dimensional boundary conditions as the model of $TiO_2$ surface (Fig. 1a). To make our models more realistic we also have studied titanium dioxide surface passivated by hydrogen atoms (Fig. 1b). For description of structural models for P and Ca doped bulk titanium we used hcp-Ti supercell consisting of 56 atoms. Taking into account the previous results obtained in [23], we have performed calculations for substitutional (*S*) and interstitial (*I*) P and Ca impurities in pure titanium. Formation energy are calculated by standard formulas

$$E_{formation} = E_{host} - [E_{host+impurity} - E_{removed} + E_{impurity}],$$

where $E_{host}$ – is the energy of system before insertion of impurity (in the case of second impurity follow to first we used the energy of the system with first impurity in the most energetically favourable position), $E_{impurity}$ and $E_{removed}$ – is the total energies per atom calculated for ground state of removed and implanted elements and $E_{host+impurity}$ – is the total energy of the system with studied impurity.

Orthophosporic acid $H_3PO_4$ crystallizes as anhydrous acid. The crystal structure has been determined using X-ray diffraction method [24]. The unit cell is monoclinic with $a = 5.78$ Å, $b = 4.84$ Å, $c = 11.65$ Å, and $\beta = 95.5$ (the space group is $P21/c$).

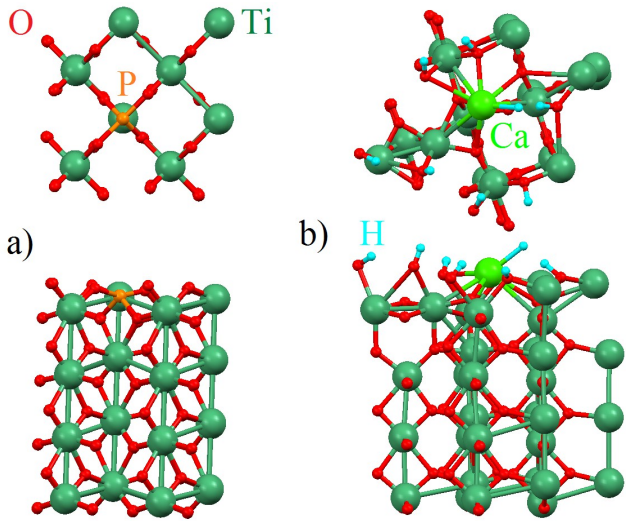

**Figure 1** Optimized atomic structure of pure $TiO_2$ (a) and passivated by hydrogen (b) slab with substitution of titanium atom (a) by phosphorous ($P_{Ti}$) and substitution of titanium atom (b) by calcium ($Ca_{Ti}$).

The crystal structure of TiP has been refined using single crystal X-ray methods [25]. The space group is P62/mmc and unit cell dimensions are $a = 3.499$ Å and $c = 11.7$ Å. Band structure calculations have been carried out within tight-binding linear muffin-tin orbital atomic spheres approximation (TB-LMTO-ASA) framework [26, 27]. The von Barth-Hedin local exchange-correlation potential has been used [28]. There is 24 **k**-points and 57 **k**-points in irreducible part of the Brillouin zone of $H_3PO_4$ and TiP, respectively.

**3 Results and discussions** The XPS survey spectra of $P^+$ or/and $Ca^+$ ion implanted samples measured at energy range of 600-0 eV are presented in Fig. 2. As seen, in spectra of $P^+$ and $Ca^+$ ion implanted *cp*-Ti besides the Ti *2s*, Ti *2p* and O *1s* lines typical for native $TiO_2$ film the additional P *2s*, P *2p* and Ca *2p* signals are detected, respectively. For double ion implanted samples Ti:P,Ca and Ti:Ca,P it is found that the implantation sequence is very important. For the P implantation followed by Ca implantation (Ti:P,Ca), less amount of P at the outermost surface is seen on the survey spectrum and can be detected only under measurements of highly-energy resolved XPS P *2p* spectra at longer time of exposure (Fig. 3a), as could be expected from the results of single P ion implantation (Ti:P). The obtained XPS survey spectrum is found to be very similar to Ca implantation (Ti:Ca) which can be explained by the fact that in this case Ca is concentrated near the surface, whereas P is implanted into the deeper regions. If the implantation is performed in the opposite order (Ti:Ca,P) the chemical surface composition changes: P is now located near the surface while Ca is implanted into the deeper layers. The XPS Ca *2p* spectrum for (Ti:Ca,P) is also detected only under longer time of exposure (Fig. 3b). The surface composition of all samples estimated from XPS survey spectra is given in Table 1. These results are in a full agreement with paper of Wieser et al. [29] where AES depth profiles of ion implanted Ti:Ca,P and Ti:P,Ca are measured and it is shown that the second implant is shifted to the surface.

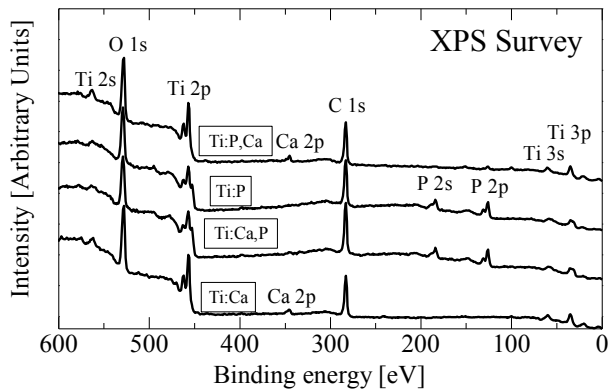

**Figure 2** XPS survey spectra of $P^+$ and $Ca^+$ ion implanted *cp*-Ti

These findings are in a good agreement with results of XRD-measurements summarized in Table 2. One can see that volume fraction of $TiO_2$ oxide on the surface of *cp*-Ti is of 6.6-10.3 vol.% and volume fraction of secondary phase CaO is higher for Ti:Ca (0.9 %) and Ti:P,Ca (0.6 %) than for Ti:Ca,P (0.2 %).

The high-energy resolved P *2p*-core level spectra of Ti:Ca:P and Ti:P,Ca ion implanted *cp*-Ti (Fig. 3a) show two-peaks structure. The low energy peak centered at ~128.3 eV is very similar to that of titanium phosphide (TiP) [30,32], which can be related to formation of Ti-P chemical bonding. On the other hand, high energy peak at ~133.3 eV is more close to that of phosphate like species [14] which evidences formation of P-O chemical bonding. This conclusion is directly confirmed by the measurements of high-energy resolved XPS Ti *2p*-spectra (Fig. 3c) where Ti $2p_{3/2, 1/2}$ binding energies at 454.1 and 464.7 eV definitely show the presence of Ti-P chemical bonding [33].

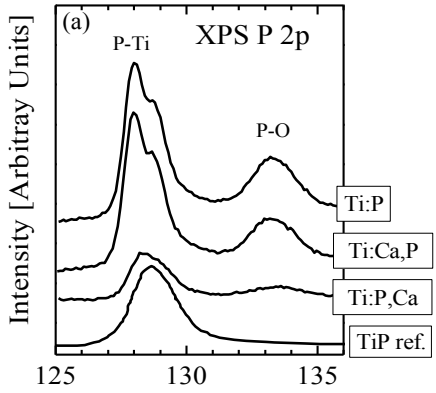
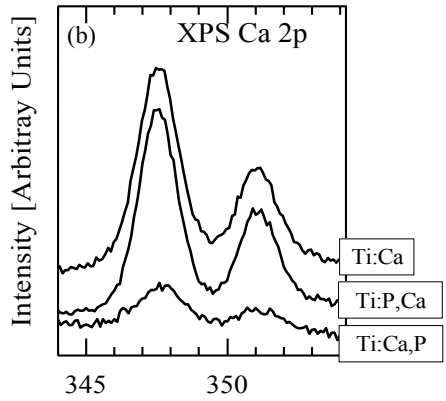
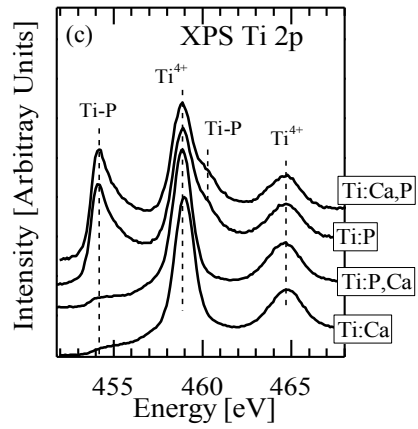

**Figure 3** XPS a) P *2p*, b) Ca *2p* and c) Ti *2p* spectra of ion implanted *cp*-Ti.

**Table 1** Surface composition of all samples (in at.%)

|  | C 1s | O 1s | P 2p | Ca 2p | Ti 2p |
|---|---|---|---|---|---|
| Ti:Ca | 47.61 | 38.94 | 0.00 | 1.13 | 12.32 |
| Ti:Ca:P | 51.31 | 27.44 | 12.12 | 0.25 | 8.88 |
| Ti:P | 48.62 | 31.33 | 11.71 | 0.00 | 8.34 |
| Ti:P,Ca | 49.12 | 36.26 | 2.15 | 1.16 | 11.31 |

**Table 2** Results of XRD measurements.

| Samp | Phase | Lattice | Volume |
|---|---|---|---|
| Ti:Ca | Ti ($P6_3/mcm$) | $a = 2.955, c =$ | rest |
| Ti:Ca,P | Ti ($P6_3/mcm$) | $a = 2.954, c =$ | rest |
| Ti:P | Ti ($P6_3/mcm$) | $a = 2.954, c =$ | rest |
| Ti:P, Ca | Ti ($P6_3/mcm$) | $a = 2.954, c =$ | rest |

For all samples, the main contribution to the formation of XPS Ti 2p-spectra comes from tetravalent titanium ($Ti^{4+}$) due to formation of Ti-O bonds similar to those in $TiO_2$ or $CaTiO_3$. XPS Ca $2p_{3/2,1/2}$ spectra of Ti:Ca and Ti:P,Ca samples have binding energies at 347.6 and 351.1 eV (Fig. 3b) which correspond to divalent Ca ions ($Ca^{2+}$) [34-35]. It can be related to formation of CaO, $CaTiO_3$ and $Ca(OH)_2$ for Ti:Ca and also to formation of $Ca_3(PO_4)_2$ for Ti:P,Ca. The measurements of XPS O 1s spectra (Fig. 4) show double-peaks structure. For Ti:P and Ti:Ca,P the observed peaks at ~530.3 and ~531.4 eV can be attributed to the formation of O-Ti bonds similar to those in $TiO_2$ or $CaTiO_3$ and O-P similar to bonds in phosphate group $[PO_4]^{3-}$. For Ti:Ca and Ti:P,Ca besides contribution from O-Ti bonds at ~530.3 eV the presence of (OH) group is found in XPS O 1s spectrum at ~532.6 eV. This indicates for formation of $Ca(OH)_2$ under $Ca^+$ and $P^+Ca^+$ ion implantation.

Results of the calculations of formation energies (Table 3) demonstrate significant energy gain under substitution of Ti atom by P in bulk titanium. This result is in agreement with extreme stability of titanium phosphate. For titanium dioxide thin films we can conclude that the most energetically favourable location of phosphorous impurities is the substitution not only of titanium but also of oxygen atoms at the surface layer of $TiO_2$ (see Fig. 1) passivated by hydrogen.

The presence of the –OH peaks in experimental spectra (Fig. 4) evidence for the hydrogen passivation of $TiO_2$ thin films formed on the surface of samples under investigation, and presence of both Ti-P and P-O peaks in XPS spectra (Figs. 3 and 4) are also suggest for the coexistence of cation and anion substitution by phosphorous.

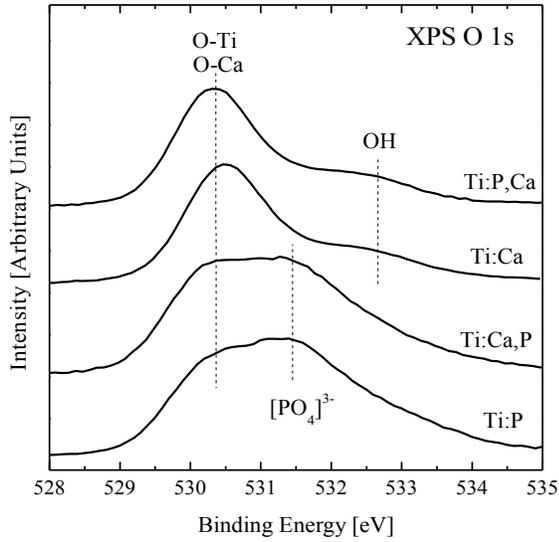

**Figure 4** XPS O 1s spectra of ion implanted *cp*-Ti.

In contrast to phosphorus, the energy required for the implantation of Ca in bulk hcp-Ti is rather high. The energies required for the substitution of Ti in $TiO_2$ is also significant. This results is in agreement with the crystal structures of real materials with an amount of Ti, Ca and O where additional atoms is required for stabilize the lattice (for example silicon in titanite – $CaTiSiO_5$).

**Table 3** Results of the calculations of formation energies.

|  |  |  |  |  |
|---|---|---|---|---|
| $TiO_2$ (hydrog | $P_{Ti}$ 3.1 | $Ca_{Ti}$ 1. | $Ca_{Ti}$ 1 | $P_{Ti}$ 2. |
| $TiO_2$ | $P_{Ti}$ 3.8 | $Ca_{Ti}$ 5. | $Ca_{Ti}$ 5 | $P_{Ti}$ 3. |
| $TiO_2$ (bulk) | $P_{Ti}$ 4.1 | $Ca_{Ti}$ 7. | $Ca_{Ti}$ 6 | $P_{Ti}$ 4. |
| Ti (bulk) | P(S) | Ca(S) |  |  |

However, cation substitution in TiO$_2$ surface is the very energetically favourable for Ca ions. The strong distortions produced by calcium ions stabilize surface of TiO$_2$ but also provide formation of Ca-O bonds for the case of substitutuion of surface oxygen by Ca ions (Fig. 1b). This result is in agreement with the presence of Ti-O, Ca-O and –OH peaks in XPS O 1s spectra (Fig. 4).

The next step is the calculation of formation energies required for the implantation of second impurity in vicinity of first one located in the most energetically favourable position. The presence of first phosphorous impurity does not provide the valuable changes in the energetics of calcium ions implantation (see Table 3) in contrast to the visible changes in the energetics of phosphorous insertion after calcium. In this case, substitution of Ti ions by P became less energetically favorable than cation substitution. These changes of the energetics of implantation provides experimentally detected absence of [PO$_4$]$^{3-}$ groups in Ti:P$^+$,Ca$^+$ (Fig. 3) in contrast with the samples where phosphorous ions are inserted first (Ti:P$^+$ and Ti:Ca$^+$,P$^+$). Significant decreasing of Ti-P peaks in XPS spectra of Ti (Fig. 3c) can be explained as result of implantation nearly all phosphorous ions in strongly distorted TiO$_2$:Ca$^+$ surface (Fig. 1b) without valuable penetration in depth and substitution of atoms in bulk titanium.

XPS valence band spectra (Fig. 5) show the essential differences between Ti:P$^+$, Ti:Ca$^+$,P$^+$ and Ti:Ca$^+$, Ti:P$^+$,Ca$^+$ systems. For Ti:P$^+$ and Ti:Ca$^+$,P$^+$ one can see two Ti *3p*-signals (with almost equal intensity) as in XPS Ti *2p*-spectra (Fig. 3c). The first one with binding energy of 37.1 eV can be attributed to the formation of Ti-O chemical bonds whereas the second one reflects the formation of Ti-P chemical bonding. In XPS VB of Ti:Ca$^+$ and Ti:P$^+$,Ca$^+$ one can observe only one strong Ti *3p* signal at ~37.1 eV belonging to Ti-O chemical bonding.

In the energy range of 28-20 eV one can see two peaks centred at ~25.3 and ~22.4 eV which are present in XPS VB spectra of all samples. The assignment of these features is rather complicated because according to the band structure calculations of Ca$_3$(PO$_4$)$_2$ and SrTi(PO$_4$)$_2$ [31, 32] in this energy range Ca *3p*, O *2s* and P *3s*-states are located. We just can mention that for Ti:P,Ca and Ti:Ca samples the signal at ~25 eV is higher than in other samples and therefore can be attributed to Ca *3p*-states due to higher content of Ca on the surface (Fig. 1 and Table 1).

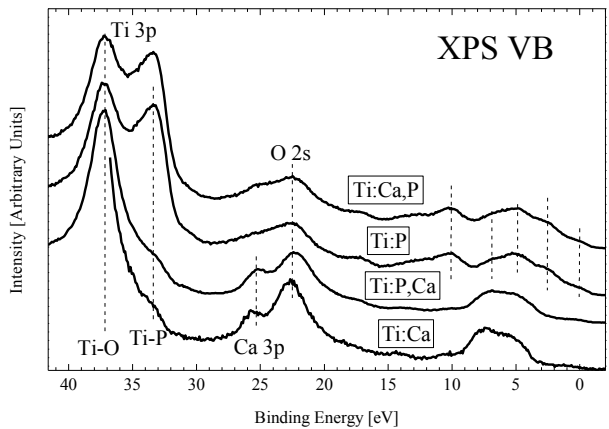

**Figure 5** XPS valence band spectra of ion implanted cp-Ti.

The valence band structure at 12-0 eV is found to be more complicated for Ti:P$^+$ and Ti:Ca$^+$,P$^+$ than for Ti:Ca$^+$ and Ti:P$^+$,Ca$^+$. To understand these differences we have compared in Fig. 6 XPS VBs of Ti:P$^+$ and Ti:Ca$^+$,P$^+$ with XPS VB spectrum of reference sample TiP [30] and results of electronic structure calculation of TiP (our data) and [PO$_4$]$^{3-}$ phosphate group of Ca$_{10}$(PO$_4$)$_6$(OH)$_2$ hydroxyapatite (HA) taken from Ref. 37. As seen, the high-energy structure *a-b* in the vicinity of the Fermi level of XPS VB of Ti:P$^+$ and Ti:Ca$^+$,P$^+$ is very similar to that of reference sample TiP [33]. According to our electronic structure calculation, it is due to Ti *3d*-P *3p* chemical bonding. The appearance of feature *e* in XPS VB of Ti:P$^+$ and Ti:Ca$^+$,P$^+$ can be attributed to Ti *4p*-P *3s* chemical interaction. The fine structure *c-f* of XPS VB is close to the total density of occupied electronic states of [PO$_4$]$^{3-}$ phosphate group of hydroxyapatite [37].

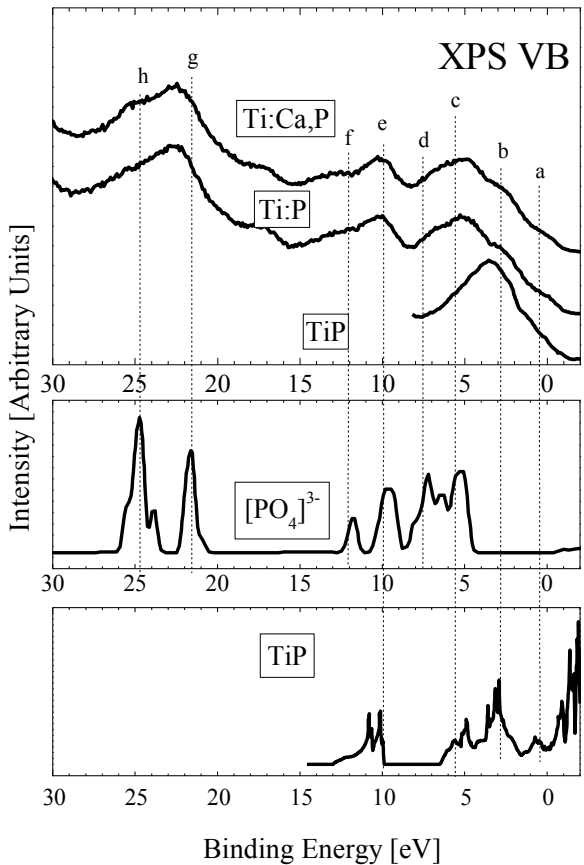

**Figure 6** Comparison of XPS VB of Ti:P and Ti:Ca,P with DOS of TiP and $[PO_4]^{3-}$ phosphate group of HA.

Our band structure calculation of $H_3PO_4$ shows that *h-g* features of XPS VBs are due to contribution of P *3s* and O *2s*-states within $[PO_4]^{3-}$ phosphate group. Basing on the comparison of XPS VBs with spectra of reference samples and electronic structure calculations we can conclude that $P^+$ and $Ca^+P^+$ ion implantation induces nucleation of TiP and $[PO_4]^{3-}$ species in native $TiO_2$ layer of *cp*-Ti.

The formation of TiP and $[PO_4]^{3-}$ species is quite favorable for biocompatibility $P^+$ and $Ca^+P^+$ ion implanted *cp*-Ti. According to Ref. 38, Ti-P has favorable biocompatibility and bone affinity, which suggest that implants with Ti-P surface may enhance osseointegration. The formation of $[PO_4]^{3-}$ species induce bioactive sites for subsequent formation of hydroxyapatite.

The Ca-ion incorporation to Ti oxide layer increases the bone-implant contact and enhances osseointegration as suggested in [39]. The same was proposed by Hanawa [39] for Ca-implanted Ti surfaces. According to these authors, the additional $Ca^{2+}$ at the surface surroundings raises the local pH which favors HA nucleation. As shown in Refs. [40-42] the dissociation of hydroxyl groups from the titania surface in the body fluid environment triggers the primary stage of HA nucleation.

## 4 Conclusions

To conclude we have performed the surface modification of commercially pure titanium (*cp*-Ti) with help of $P^+$, $Ca^+$, $P^+Ca^+$ and $Ca^+P^+$ ion implantation (E=30 keV, D=1x10$^{17}$ cm$^{-2}$). XPS measurements of core levels show that for double ion implantation the implantation sequence is very important: for $P^+$ implantation followed by $Ca^+$ ($P^+Ca^+$) calcium is concentrated near the surface whereas phosphorus is implanted into the deeper layers and opposite situation is observed for ($Ca^+P^+$) implantation. For characterization of ion implanted *cp*-Ti at first time the measurements of XPS valence bands were performed which were compared with specially performed density functional theory calculations allowing to receive a full information about electronic structure of surface of these materials. In accordance with XPS VBs measurements the density functional theory calculations show that for $P^+$ and $Ca^+P^+$ implantation the formation energies for cation (P-Ti) and anion (P-O) substitution are comparable which induce the creation of TiP and $[PO_4]^{3-}$ species. For $Ca^+$ and $P^+Ca^+$ implantation only cation $Ca^+$-$Ti^{4+}$ substitution takes place. The conversion of implanted ions to $Ca^{2+}$, TiP and $[PO_4]^{3-}$ species is favorable for biocompatibility and, as shown in Refs [39-42], assists to initiate the bioactive sites for subsequent formation of hydroxyapatite.

**Acknowledgements** We acknowledge support of the Russian Scientific Foundation (Project 14-22-00004).

## References

[1] S. Nag, R. Banerjee R Fundamentals for Medical Implants, ASM Handbook Materials for Medical Devices, *Vol. 23*, (Ed. By R. Narayan), **2012**, pp. 6-17
[2] D. F. Williams in Biocompatibility of clinical implant materials (ed. By D. F. Williams) Boca Raton: CRC Press, **1981**.
[3] B. D. Ratner in Titanium in Medicine (Eds. D.M. Brunette, P. Tengvall, M. Textor, P. Thomsen) Berlin and Heidelberg: Springer-Verlag, **2001**, pp. 1–12
[4] P. Thomsen, C. Larsson, L. E. Ericson, L. Sennerby, J. Lausmaa, B. Kasmo, *J. Mater. Sci. Mater. Med.* **1998**, *8*, 653.
[5] R. Narayanan, S. K. Seshadri, T. Y. Kwon, K. H. Kim, *J. Biomed. Mater. Res.* **2008**, *85B*, 279.
[6] R. Tapash, R. Rautray, R. Narayanan, K.-H. Kim, *Progr. Mater. Sci.* **2011**, *56*, 1137.
[7] J. L. Ong, L. C. Lucus, W. R. Lacefield, E. D. Rigney, *Biomaterials* **1992**, *13*, 249.
[8] I. M. O. Kangasniemi, C. C. P. M. Verheyen, E. A. van der Velde, K. de Goot, *J. Biomed. Mater. Res.* **1994**, *28*, 563.
[9] P. Sioshansi, E. J. Tobin, *Surf. Coat. Technol.* **1996**, *83*, 175.
[10] H. Herman, K. J. Hirvonen, Treatise on materials science and technology, ion implantation. **1980** New York: Academic Press.
[11] T. Hanawa, H. Ukai, K. Murakami, *J. Electron Spectrosc. Relat. Phenom.* **1993**, *63*, 347.
[12] T. Hanawa, Y. Kamiura, S. Yamamoto, T. Kohgo, A. Amemiya, H. Ukai, K. Murakami, K. Asaoka, *J. Biomed. Mater. Res.* **1997**, *36*, 131.
[13] H. Baumann, K. Bethge, G. Bilger, D. Jones, I. Symietz, *Nucl. Instr. Meth. B* **2002**, *196*, 286.
[14] S. Krischok, C. Blank, M. Engel, R. Gutt, G. Ecke, J. Schawohl, L. Spies, F. Schrempel, G. Hildebrand, K. Liefeith, *Surf. Sci.* **2007**, *601*, 3856.
[15] D. Krupa, J. Baszkiewicz, J. A. Kozubowski, M. Lewandowska-Szumieł, A. Barcz, J. W. Sobczak, A. Bilinski, A. Rajchel, *Bio-Med. Mater. Engin.* **2004**, *14*, 525.
[16] M. T. Pham, H. Reuther, W. Maitz, R. Mueller, G. Steiner, S. Oswald, I. Zyganov, *J. Mater. Sci.: Mater. Medic.* **2000**, *1*, 383.
[17] M. T. Pham, W. Matz, H. Reuther, E. Richter, G. Steiner, S. Oswald, *Surf. Sci. Technol.* **1999**, *111*, 103.
[18] Y. Xie, X. Liu, P. K. Chu, C. Ding, C *Surf. Sci.* **2006**, *600*, 651.
[19] S. Tougard, *Solid State Communs.* **1987**, *61*, 547.
[20] N. N. Bao, H. M. Fan, J. Ding, J. B. Yi, *J. Appl. Phys.* **2011**, *109*, 07C302.
[21] G. S. Chang, E. Z. Kurmaev, D. W. Boukhvalov, L. D. Finkelstein, A. Moewes, H. Bieber, S. Colis, A. Dinia, *J. Phys.: Condens. Matter* **2009**, *21*, 056002.
[22] J. P. Perdew, K. Burke, M. Ernzerhof, *Phys. Rev. Lett.* **1996**, *77*, 3865.
[23] A. Metz, S. Frota-Pessôa, J. Karpoor, D. Riegel, W. D. Brewer, R. Zeller, *Phys. Rev. Lett.* **1993**, *71*, 3525.
[24] S. Furberg, *Acta Chem. Scand.* **1954**, *8*, 532.
[25] P.-O. Snell, *Acta Chem. Scand.* **1967**, *21*, 1773.
[26] R. O. Jones, O. Gunnarsson, *Rev. Mod. Phys.* **1989**, *61*, 689.
[27] O. K. Andersen, O. Jepsen, *Phys. Rev. Lett.* **1984**, *53*, 2571.
[28] U. von Barth, L. Hedin, *J. Phys. C* **1972**, *5*, 1629.
[29] E. Wieser, I. Tsyganov, W. Matz, H. Reuther, S. Oswald, T. Pham, E. Richter, Surf. Coat. Technol. 111 (1999) 103.
[30] S. Baunack, S. Oswald, D. Scharnweber, *Surf. Interface Anal.* **1998**, *26*, 471.
[31] L. Liang, P. Rulis, W. Y. Ching *Acta Biomaterialia* **2010**, 6, 3763
[32] D. Zhao, H. Zhang, Z. Xie, W. L. Zhang, S. L. Yang, W. D. Cheng *Dalton Trans.*, **2009**, 5310.
[33] C. E. Myers, H. F. Franzen, J. W. Anderegg, *Inorg. Chem.* **1985**, *24*, 1822.

[34] C. C. Chusuei, D. W. Goodman, *Anal. Chem.* **1999**, *71*, 149.
[35] V. V. Atuchin, V. G. Kesler, N. V. Pervukhina, Z. Zhang, *J. Electr. Spectr. Relat. Phenom.* **2006**, *152*, 18.
[36] N. Ohtsu, K. Sato, K. Asami, T. Hanawa, *J. Mater. Sci.: Mater. Med.* **2007**, *18*, 1009.
[37] L. Calderín, M. J. Stott, A. Rubio, *Phys. Rev. B* **2003**, *67*, 134106.
[38] K. J. Lee, C. S. Kim, H. S. Kim, C. Y. Yum, B. O. Kim, K. Y. Han, *J. Korea. Acad. Periodontol.* **1997**, *27*, 329.
[39] J.-W. Park, K.-B. Park, J.-Y. Suh, J-Y 2007 *Biomaterials* **28** 3306
[40] T. Hanawa, *Mater. Sci. Eng. A* **1999**, *267*, 260.
[41] J.-H. Lee, S.-E. Kim, Y.-J. Kim, C.-S. Chi, H.-J. Oh, *Mater. Chem. Phys.* **2006**, *98*, 39.
[42] W. Simka, A. Iwaniak, G. Nawrat, A. Maciej, J. Michalska, K. Radwanski, J. Gazdowicz, *Electrochim. Acta* **2009**, *54*, 6983.